\documentclass[a4paper]{article}

\usepackage{INTERSPEECH2019}
\usepackage{amssymb}
\usepackage{booktabs}
\usepackage{multirow}
\usepackage{graphics}
\usepackage{float,threeparttable}

\usepackage{commath}
\usepackage{mathtools}
\usepackage{here}
\usepackage{url}
\usepackage{cite}
\usepackage{array,etoolbox}
\usepackage{caption}
\preto\tabular{\setcounter{magicrownumbers}{0}}
\newcounter{magicrownumbers}

\usepackage{stackengine}
\stackMath

\title{An Investigation of End-to-End Multichannel Speech Recognition for Reverberant and
Mismatch Conditions}
\name{Aswin Shanmugam Subramanian$^1$, Xiaofei Wang$^1$, Shinji Watanabe$^1$\\
Toru Taniguchi$^2$, Dung Tran$^2$, Yuya Fujita$^2$}
\address{
  $^1$Center for Language and Speech Processing,
            Johns Hopkins University, Baltimore, MD, USA\\
  $^2$Yahoo Japan Corporation, Tokyo, Japan}
\email{aswin@jhu.edu}
\email{\{aswin, xiaofeiwang, shinjiw\}@jhu.edu, \{ttaniguc, tdung, yuyfujit\}@yahoo-corp.jp}
\begin{document}

\maketitle
\begin{abstract}
Sequence-to-sequence (S2S) modeling is becoming a popular paradigm for automatic speech recognition (ASR) because of its ability to jointly optimize all the conventional ASR components in an end-to-end (E2E) fashion. This report investigates the ability of E2E ASR from standard close-talk to far-field applications by encompassing entire multichannel speech enhancement and ASR components within the S2S model. There have been previous studies on jointly optimizing neural beamforming alongside E2E ASR for denoising. It is clear from both recent challenge outcomes and successful products that far-field systems would be incomplete without solving both denoising and dereverberation simultaneously. This report uses a recently developed architecture for far-field ASR by composing neural extensions of dereverberation and beamforming modules with the S2S ASR module as a single differentiable neural network and also clearly defining the role of each subnetwork. The original implementation of this
architecture was successfully applied to the noisy speech recognition task (CHiME-4), while we
applied this implementation to noisy reverberant tasks (DIRHA and REVERB). Our investigation
shows that the method achieves better performance than conventional pipeline methods on the DIRHA English dataset and comparable performance on the REVERB dataset. It also has additional advantages of being neither iterative nor requiring parallel noisy and clean speech data.
\end{abstract}

\section{Introduction}
Sequence-to-sequence (S2S) neural network models for automatic speech recognition (ASR) \cite{bahdanau2016end} are rapidly gaining a lot of attention and popularity because of their property to jointly optimize all the conventional ASR components in an end-to-end (E2E) fashion.
It is seen as a competitive alternate to state-of-the-art hidden Markov model (HMM)-deep neural network (DNN) based hybrid automatic speech recognition (ASR) systems \cite{hinton2012deep} as it has achieved comparable performance on tasks with a very large amount of training data \cite{amodei2016deep,google_sor}.
The legacy hybrid ASR system has multiple components optimized independently and works in a soft pipeline fashion where the (probabilistic) output of the preceding component is fed as an input to the succeeding component. 
On the other hand, the E2E model just composes of a single network, which is trained to map a sequence of speech features directly to a text sequence, by optimizing all the different components in the ASR pipeline jointly. 

This report investigates the above joint optimization ability of E2E ASR from standard close-talk to far-field applications by encompassing entire multichannel speech enhancement and ASR components within the S2S model.
Far-field ASR systems often utilize input from multiple microphones and have frontend enhancement components to handle distortions caused by both noise and reverberation \cite{rasr_jli}. 
Outcomes of recent challenges like REVERB \cite{kinoshita2016summary} and CHiME-5 \cite{chime5is} show that both denoising and dereverberation components are indispensable for handling far-field speech. 
Typically for hybrid ASR, a multichannel dereverberation component followed by a beamforming component is used as an additional pipeline \cite{drude2018integrating}. 
Instead of using these techniques as a pipeline for E2E models or simply extending E2E models to allow multichannel speech features \cite{braun2018multi,wang2018stream}, it is straightforward to include carefully designed sub-networks for beamforming and dereverberation within the E2E model to take advantage of the fact that they can be jointly trained.

Currently, neural beamforming techniques for denoising \cite{nn-gev, erdogan2016improved} have given state-of-the art results in robust ASR tasks like the CHiME-4 challenge \cite{du2016ustc,menne2016rwth,vincent2017analysis, Chen2018}. 
In these techniques, a neural network is employed to estimate speech and noise masks, which in turn are used to compute the power spectral density (PSD) matrices needed to estimate the beamforming filter. 
The neural network is often trained with simulated data and the ground truth mask as the target. 
A neural beamforming mechanism as a differentiable component was proposed in \cite{Ochiai2017MultichannelES,ochiai2017unified} to allow the joint optimization of multichannel speech recognition and enhancement within the E2E system only based on the ASR objective.
The masks are made as latent variables in the end-to-end training. Hence, parallel clean and noisy speech data are not needed in this approach. 
Joint training of a neural beamformer with a hybrid ASR acoustic model is also proposed in \cite{xiao2016deep,google_abf,beamnet,menne2018speaker,minhua2019frequency}, although these require frame-level alignments, while the E2E system does not.

Weighted prediction error (WPE) \cite{wpe, yoshioka2012generalization} is a technique based on variance normalized long term linear prediction popularly used for dereverberation of  \textit{wet} (reverberant) signals. 
It has been very effective in successful commercial products like Google Home \cite{li2017acoustic}. This technique requires an estimate of the time-varying variance of the desired \textit{dry} signal and hence the conventional WPE method is iterative. A non-iterative method DNN-WPE was proposed in \cite{dnnwpe,heymann2018frame} where a neural network was trained to estimate the magnitude spectrum of the desired signal from the observed signal's magnitude spectrum. 
We can introduce a mask as a hidden state vector similar to \cite{Ochiai2017MultichannelES} for estimation of the magnitude spectrum of the desired signal without parallel data. 
The WPE filtering solution being differentiable, we can train this also in an E2E framework and optimize it only based on the ASR objective.

We use the implementation developed by NTT which has neural extensions of WPE and MVDR with E2E ASR \footnote{https://github.com/espnet/espnet/pull/596}. The original implementation was applied on noisy data CHiME4 without including the dereverberation component. We investigate to jointly train both WPE based dereverberation and minimum variance distortionless response (MVDR) based beamforming along with ASR using noisy and reverberant data and also test it on mismatch conditions. Other major differences from the original implementation are as follows: (1) we jointly train both dereveberation and beamforming by passing the input through both subnetworks while the original implementation only passes through one of them, (2) we also investigate applying a speech activity detection type mask in the beamforming subnetwork and different activations for the mask in the dereverberation subnetwork. The parameters to be estimated by the neural network for the front-end are channel-independent masks. This makes the trained system to generalize for input signals with arbitrary number and order of channels like \cite{Ochiai2017MultichannelES}. 

\section{Multi-channel end-to-end ASR}
\subsection{Dereverberation subnetwork}
\label{sec:dereverb}
 This section explains WPE based dereverberation method \cite{wpe, yoshioka2012generalization,kinoshita2016summary}, which cancels late reverberations using variance normalized delayed linear prediction (NDLP). WPE estimates the desired $M$-channel (dereverberated) signal $\mathbf{d}(t, b) \in \mathbb{C}^M$ in the short-time Fourier transform (STFT) domain at time frame $t$, frequency bin $b$ using the following vector-form equation:
\begin{equation}
\mathbf{d}(t, b) = \mathbf{y}(t, b) - \mathbf{G}^{\textrm{H}}(b) \mathbf{\tilde{y}}(t-\Delta, b),
\label{eq:desire}
\end{equation}
$\mathbf{y}(t, b) \in \mathbb{C}^{M}$ is the observed multichannel signal in the STFT domain, $\Delta$ is the prediction delay.
$\mathbf{G}^{\textrm{H}}(b) \in \mathbb{C} ^{ML\times M}$ and $\mathbf{\tilde{y}}(t-\Delta, b) \in \mathbb{C} ^{ML}$ are the stacked representations of the prediction filter coefficients and the delayed multichannel observations with the filter order $L$, respectively.
$\mbox{}^{\textrm{H}}$ denotes the conjugate transpose.

WPE assumes the desired signal $\mathbf{d}(t, b)$ in Eq.~\eqref{eq:desire} is a realization of a zero-mean complex Gaussian $\mathcal{N}^{\text{c}}(\cdot)$ with an unknown channel independent time-varying variance $\lambda(t, b) \in \mathbb{R} _{>0}$, as follows:
\begin{equation}
p(\mathbf{d}(t, b); \lambda(t, b)) =  \mathcal{N}^{\text{c}}(\mathbf{d}(t, b); \mathbf{0},\lambda(t, b)\mathbf{I}). 
\end{equation}
The prediction filter $\mathbf{G}(b)$ in Eq.~\eqref{eq:desire} obtained based on maximum likelihood estimation yields the following iterative solution with the previously estimated desired signal $\bar{d}(t, b, m)$:
\begin{align}
\lambda(t, b) &=  \frac{1}{M} \sum\limits_{m} \abs{\bar{d}(t, b, m)}^{2}, \label{eq:step1_1} \\
\mathbf{R}(b) &=  \sum\limits_{t} \frac{\mathbf{\tilde{y}}(t-\Delta, b)\mathbf{\tilde{y}}^{\textrm{H}}(t-\Delta, b)}{\lambda(t, b)} \label{eq:step1_2}, \\
\mathbf{P}(b) &=  \sum\limits_{t} \frac{\mathbf{\tilde{y}}(t-\Delta, b)\mathbf{y}^{\textrm{H}}(t, b)}{\lambda(t, b)} \in \mathbb{C} ^{ML\times M}, \label{eq:step2_1}\\
\mathbf{G}(b) &=  \mathbf{R}(b)^{-1}\mathbf{P}(b) \in \mathbb{C} ^{ML\times M}, \label{eq:step2_2}
\end{align}
where $m$ is the channel index and $\mathbf{R}(b) \in \mathbb{C} ^{ML\times ML}$ is the correlation matrix. 
In conventional WPE, the estimated desired signal $\bar{d}(t, b, m)$ in Eq.~\eqref{eq:step1_1} is initialized with the observed signal $y(t, b, m)$ to estimate the variance $\lambda(t, b)$ in the first iteration. 
This iterative process makes this algorithm slow and also loses online processing capabilities.

Instead, \cite{dnnwpe} uses a DNN to estimate the magnitude spectrum $|\bar{d}(t, b, m)|$ in Eq.~\eqref{eq:step1_1} from the magnitude spectrum of the observed signals $\abs{y(:,:,m)}$ for every channel $m$\footnote{We use the notation $f(:)$ to denote all elements. For example, $y(:,:,m)$ denotes the observation STFT signal of a channel $m$ for all frames and frequency bins.}.
This network gives a good estimate of the variance $\lambda(t, b)$, and it was shown that the performance obtained with one-shot filter estimation with Eqs.~\eqref{eq:step1_2}--\eqref{eq:step2_2} can match that of WPE without iterations. 
This method is called DNN-WPE. 
The drawback in this method is we need to simulate parallel data to train the DNN.

We use this WPE-based dereverberation as a sub-network of our E2E framework (described in Section \ref{sec:dftnet}).
This sub-network processing (defined as the operation WPE$(\cdot)$) is based on the sequence of the filter estimation steps based on Eqs.\eqref{eq:step1_1}--\eqref{eq:step2_2}, and the final dereverberation based on Eq.~\eqref{eq:desire}.
As all the operations are easily differentiable, we can incorporate it into a computational graph for joint training. 
In our joint training approach, we propose to estimate the desired power spectrum $\abs{d(t, b, m)}^{2}$ via the following masking network $\text{MaskNet}_{\text{D}}(\cdot)$ that produces a mask $w(t, b, m)\in [0, 1]$:
\begin{align}
w(:,:,m) & = \text{MaskNet}_{\text{D}}(y(:,:,m)) \label{eq:dmask_1}, \\
\abs{d(t, b, m)}^{2} & = w(t, b, m) \abs{y(t, b, m)}^{2} \label{eq:dmask_2},
\end{align}
Since the domain of the mask is bounded within $[0, 1]$, it is easily estimated from a neural network compared with the direct prediction of the desired power/magnitude spectrum.

\begin{figure}[t]
  \centering
  \includegraphics[width=\linewidth]{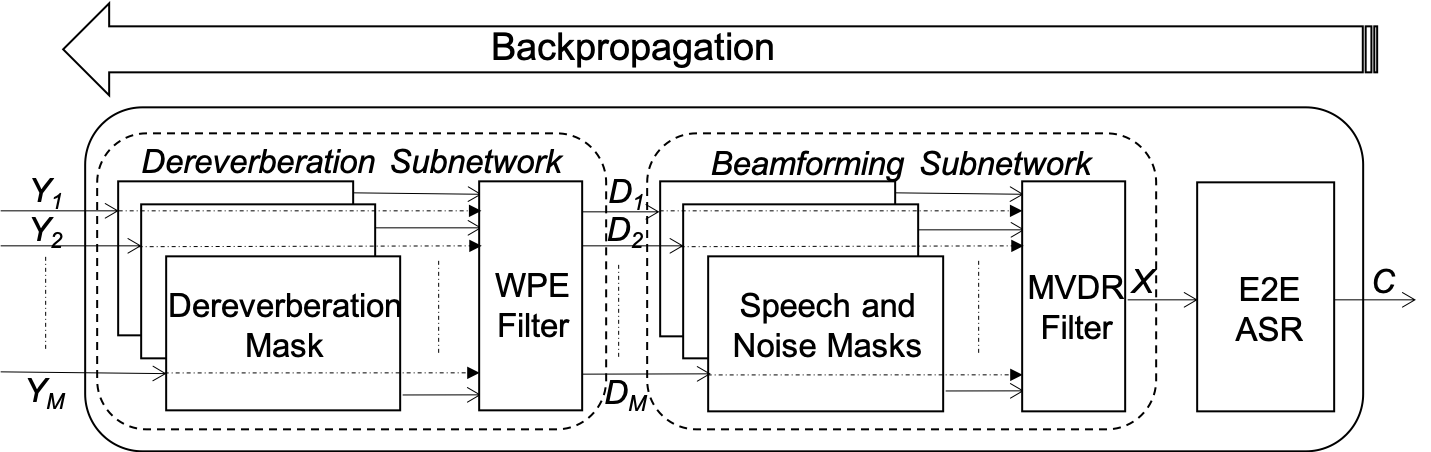}
  \caption{E2E Multichannel ASR architecture.}
  \label{fig:dftnet}
\end{figure}


\subsection{Beamforming subnetwork}
We use a beamforming subnetwork similar to the one proposed in \cite{Ochiai2017MultichannelES}. 
Like the dereverberation subnetwork, another set of two masking networks $\text{MaskNet}_{\text{S}}$ and are $\text{MaskNet}_{\text{N}}$ are used to produce the speech mask $w _{\text{S}} (t, b, m) \in [0, 1]$ and the noise mask $w _{\text{N}} (t, b, m) \in [0, 1]$ given the output $d(t, b, m)$ from $\text{WPE}(\cdot)$, as follows:
\begin{align}
    w _{v} (:, :, m) = \text{MaskNet}_{v}(d(:, :, m)) \ \textrm{where} \  v \in \{\text{S}, \text{N}\} \label{eq:mask_b}.
\end{align}
These masks are averaged over channels (e.g., $w _{v} (t, b) = 1/M \sum _m w _{v} (t, b, m)$) and used to compute the power spectral density (PSD) matrices of speech and noise $\mathbf{\Phi}_{\text{S}}(b) \in \mathbb{C} ^{M\times M}$ and $\mathbf{\Phi}_{\text{N}}(b) \in \mathbb{C} ^{M \times M}$ at frequency bin $b$ as follows:
\begin{equation}
\mathbf{\Phi}_{v} (b) = \sum\limits_{t=1}^T w_v(t,b)\mathbf{d}(t, b)\mathbf{d}^\textrm{H}(t, b) \  \textrm{where} \  v \in \{\text{S}, \text{N}\},
\label{psd}
\end{equation}
From these PSD matrices, the $M$-dimensional complex MVDR beamforming filter $\mathbf{f}_{\textrm{MVDR}} (b) \in \mathbb{C} ^M$ is estimated by solving the following optimization problem:

\begin{equation}
\mathbf{f}_{\textrm{MVDR}} (b) = \frac{\mathbf{\Phi}_{\text{N}} (b)^{-1}\mathbf{\Phi}_{\text{S}} (b)}{\text{Tr}(\mathbf{\Phi}_{\text{N}} (b)^{-1}\mathbf{\Phi}_{\text{S}} (b))} \mathbf{u} 
\label{estimate_bf}
\end{equation}
where $\mathbf{u} \in \{0, 1\} ^{M}$ is a one-hot vector to choose a reference microphone and the beamformer estimates the speech image at the reference microphone. 
$\text{Tr}(\cdot)$ denotes the trace operation.
We use the MVDR formulation based on reference selection (Eq.~\eqref{estimate_bf}) given in \cite{mvdr_souden} instead of the widely-used steering vector estimation based formulation \cite{7805139} to make the operation more easily differentiable.   
Note that all the masking networks in beamforming (and dereverberation) are trained without any signal-level supervision but with the ASR objective. 

Once we obtain the beamforming filter $\mathbf{f}_{\textrm{MVDR}} (b)$, we can perform speech denoising to obtain an enhanced STFT signal $x(t, b) \in \mathbb{C}$ as follows:
\begin{equation}
 x (t, b) = \mathbf{f}_{\textrm{MVDR}}^\textrm{H} (b) \mathbf{d}(t, b).
\label{apply_bf}
\end{equation}
Similar to the WPE operation, we define this MVDR filter estimation using Eqs.~\eqref{psd} and \eqref{estimate_bf} along with the denoising equation \eqref{apply_bf} as MVDR$(\cdot)$.

\begin{table*}[htbp]
  \centering
    \caption{WER (\%) on REVERB and DIRHA-WSJ (LA array)  evaluation sets comparing the performance of pipeline \& E2E frontend techniques.}
    \resizebox{\textwidth}{!}{%
    \begin{tabular}{|c|c|c|c|c|c|c|c|c|c|c|c|c|c|}
    \toprule
    \toprule
    \multicolumn{1}{|c|}{\multirow{3}[6]{*}{\textit{\textbf{Method}}}} & \multicolumn{1}{c|}{\multirow{3}[6]{*}{\textit{\textbf{Dereverberation}}}} & \multicolumn{3}{c|}{\textit{\textbf{Beamformer}}} & \multicolumn{6}{c|}{\textit{\textbf{REVERB Simulated}}} & \multicolumn{2}{c|}{\textit{\textbf{REVERB Real}}} & \multicolumn{1}{c|}{\multirow{3}[6]{*}{\textit{\textbf{DIRHA LA}}}} \\
\cmidrule{3-13}          &       & \multicolumn{1}{c|}{\multirow{2}[4]{*}{\textit{\textbf{Method}}}} & \multicolumn{1}{c|}{\multirow{2}[4]{*}{\textit{\textbf{Reference}}}} & \multicolumn{1}{c|}{\textit{\textbf{Mask}}} & \multicolumn{2}{c|}{\textit{\textbf{Room 1}}} & \multicolumn{2}{c|}{\textit{\textbf{Room 2}}} & \multicolumn{2}{c|}{\textit{\textbf{Room 3}}} & \multicolumn{2}{c|}{\textit{\textbf{Room 1}}} &  \\
\cmidrule{6-13}          &       &       &       & \multicolumn{1}{c|}{\textit{\textbf{Type}}} & \textit{\textbf{Near}} & \textit{\textbf{Far}} & \textit{\textbf{Near}} & \textit{\textbf{Far}} & \textit{\textbf{Near}} & \textit{\textbf{Far}} & \textit{\textbf{Near}} & \textit{\textbf{Far}} &  \\
    \midrule
    Challenge baseline \cite{kinoshita2016summary}    & -     & -     & -   &-  & 16.2 & 18.7   & 20.5   & 32.5  & 24.8   & 38.9  & 50.1  & 47.6  & - \\
    E2E Baseline & -     & -     & -     & -     & \textbf{5.4} & 7.1   & 7.6   & 12.9  & 9.7   & 16.1  & 23.9  & 26.8  & 55.3 \\
    \midrule
    \multirow{5}[2]{*}{Pipeline} & WPE   & -     & -     & -     & 6.0   & 6.6   & 7.1   & 9.8   & 8.0   & 11.2  & 17.7  & 18.4  & 42.3 \\
          & DNN-WPE & -     & -     & -     & 5.7   & 6.0   & 7.5   & 9.3   & 7.8   & 10.1  & 16.4  & 18.5  & 41.3 \\
          & -     & BeamformIt & X-Corr & -     & 5.8   & 6.1   & \textbf{5.8} & 8.5   & 6.9   & 10.2  & 14.6  & 16.1  & 39.2 \\
          & WPE   & BeamformIt & X-Corr  & -     & 6.6   & 5.9   & 6.1   & 7.0   & 6.8   & 8.2   & 11.3  & 11.9  & \textbf{30.7} \\
          & DNN-WPE & BeamformIt & X-Corr  & -     & 6.3   & \textbf{5.8} & 6.4   & \textbf{6.8} & \textbf{6.6} & \textbf{7.7} & \textbf{11.0} & \textbf{10.8*} & 31.3 \\
    \midrule
    \multirow{6}[2]{*}{E2E} & WPE   & -     & -     & -     & 6.3   & 6.7   & 6.7   & 8.9   & 7.4   & 10.6  & 17.0  & 19.8  & 42.3 \\
          & -     & MVDR  & Ch 2   & TF    & 5.7   & 6.1   & 5.6   & 8.2   & 6.2   & 10.2  & 12.6  & 17.3  & 42.3 \\
          & -     & MVDR  & Ch 2   & SAD   & 7.2   & 7.2   & 6.4   & 8.6   & 7.1   & 12.1  & 16.0  & 20.5  & 45.3 \\
          & WPE   & MVDR  & Ch 2   & TF    & \textbf{5.5} & \textbf{5.7*} & \textbf{5.3*} & \textbf{6.6*} & 6.5   & \textbf{7.6*} & 10.7  & 13.7  & 35.4 \\
          & WPE   & MVDR  & Ch 2   & SAD   & 8.3   & 7.8   & 6.9   & 7.0   & 7.6   & 8.6   & 10.8  & 13.9  & 31.6 \\
          & WPE   & MVDR  & Attention & SAD   & 6.4   & 6.3   & 5.9   & 6.8   & \textbf{6.3*} & \textbf{7.6*} & \textbf{8.7*} & \textbf{12.4} & \textbf{29.1*} \\
           \midrule
    Tachioka et. al. \cite{tachioka2014dual}      &  Spectral subtraction    &  Delay-sum  &-   & - &	5.0	&5.6&	5.6&	8.2	&5.7	&10.5&	16.9&	20.3 &- \\
    Alam et. al. \cite{alam2014use} &  Iterative deconvolution    &  - &- & - &	6.7	&7.3&	8.0&	11.1	& 8.1	&12.1&	21.4&	22.0 &- \\
    Wang et. al. \cite{wang2018stream} & - & - & - & - & - & - & - & - & - & - & - & - & 35.1 \\
    \bottomrule
    \bottomrule
    \end{tabular}}
  \label{tab:WER}%
  \vspace*{-4mm}
\end{table*}%

\subsection{Joint dereverberation \& beamforming}
\label{sec:dftnet}
The beamforming subnetwork is placed after the dereverberation subnetwork. 
The output of the beamforming network goes to the ASR. 
This whole network is trained solely based on the ASR objective. The architecture is shown in Figure \ref{fig:dftnet}. 

We summarize the three stages in terms of operations defined before as: 
\begin{itemize}
\item \textbf{Dereverberation}: Eqs.~\eqref{eq:dmask_1} $\rightarrow$ \eqref{eq:dmask_2} $\rightarrow$  \eqref{eq:step1_1}--\eqref{eq:step2_2} $\rightarrow$ \eqref{eq:desire}
\begin{align}
    w(:, :, m) & = \text{MaskNet}_{\text{D}}(y(:, :, m)). \\
    D & = \text{WPE}(W, Y).
\end{align}
\item \textbf{Beamforming}: Eqs.~\eqref{eq:mask_b} $\rightarrow$ \eqref{psd} $\rightarrow$ \eqref{estimate_bf} $\rightarrow$ \eqref{apply_bf}
\begin{align}
    w_{\text{S}}(:, :, m) & = \text{MaskNet}_{\text{S}}(d(:, :, m)). \\
    w_{\text{N}}(:, :, m) & = \text{MaskNet}_{\text{N}}(d(:, :, m)). \\
    X & = \text{MVDR}(W_{\text{S}}, W_{\text{N}}, D).
\end{align}
\item \textbf{Feature extraction \& recognition}:
\begin{align}
    F & = \text{MVN}(\text{Log}(\text{MelFilterbank}(\abs{X})))\\
    C & = \text{ASR}(F).
\end{align}
\end{itemize}
where $W$, $Y$, and $D$ denote the mask, observation and dereverberated STFT signals for all frames, frequency bins, and channels, respectively.
$W_{\text{S}}$ and $W_{\text{N}}$ denote the averaged speech and noise masks over the channels and $X$ denotes the beamformed STFT signal for all frames and frequency bins. 
Log Mel Filterbank transformation is applied on the magnitude of $X$ and utterance based mean-variance normalization (MVN) is performed to produce an input that is suitable for ASR $F$.
All these operations are still differentiable.
$C = (c_1, c_2, \cdots)$ is the character sequence that represents the text output of E2E ASR ($\text{ASR}(\cdot)$).

One of the most important benefits of this architecture is that the entire network is represented as a differentiable computational graph and their parameters are jointly trained using back propagation, as shown in Figure~\ref{fig:plot}.
This architecture also clearly defines the role of each subnetwork by careful design of their architecture which makes it possible to interpret their intermediate outputs $D$ and $X$ as dereverberated and denoised signals respectively.

\section{Experiments}
\subsection{Setup}
We evaluated the effectiveness of the method described in the previous section by using the REVERB \cite{kinoshita2016summary} and DIRHA English WSJ \cite{dirha} datasets in the following way. 
\begin{itemize}
    \item \textbf{Training} - 2-channel simulation data from REVERB and clean data from wall street journal (WSJ) corpus \cite{paul1992design} (both WSJ0 and WSJ1)
    \item \textbf{Validation} - REVERB 8-channel real and simulation development sets.
        \item \textbf{Evaluation} - (1) REVERB 8-channel real and simulation evaluation sets, (2) DIRHA-WSJ 6-channel real recordings from the living room's circular ceiling array.  
\end{itemize}


A hybrid combination of connectionist temporal classification (CTC) and attention-based
encoder-decoder model \cite{watanabe2017hybrid, kim2017joint} was used for E2E speech recognition. The ESPnet toolkit \cite{espnet} was used for the E2E ASR experiments. 
The baseline E2E ASR uses the $80$-dimensional log Mel filterbank energies as the feature. The encoder consists
of two initial blocks of convolution layers followed by three output gate projected
bidirectional long short-term memory (BLSTMP) layers with 1024 units. The location based attention mechanism was used. 
The decoder consists of a single LSTM layer with 1024 units followed by a linear layer with a number of output units corresponding to the number of distinct characters.
The CTC-attention interpolation weight was fixed as ``0.5". The word based RNN language model proposed in \cite{wordlm} was used.

An open source implementation of WPE \cite{Drude2018NaraWPE} was used. 
We also implemented the DNN-WPE model \cite{dnnwpe}, and made it publicly available as an open source toolkit \footnote{https://github.com/sas91/jhu-neural-wpe}. 
The WPE filter order $L$ and the prediction delay $\Delta$, which are introduced in Section \ref{sec:dereverb}, were fixed as ``5" and ``3" respectively for all the dereverberation methods. 
The number of iterations was fixed as ``3" for WPE. 
In DNN-WPE, the architecture given in \cite{dnnwpe} was used for the neural network predicting magnitude spectrum. 
Both dereverberation subnetwork's and beamforming subnetwork's masking networks in Section \ref{sec:dftnet} consist of two BLSTMP layers followed by an additional feedforward layer. 
The dereverberation subnetwork uses clipped rectified linear unit (ReLU) with a max clamp at ``1" as the activation and the beamforming subnetwork uses sigmoid as the activation. 

We used two methods for the reference microphone selection ($\mathbf{u}$ in Eq~\eqref{estimate_bf}) in beamforming subnetwork: 1) fixing channel 2 as the reference, 2) attention based soft reference selection proposed in \cite{Ochiai2017MultichannelES} that includes reference selection inside the network using the state vectors of the masking network and the speech PSD matrix.  BeamformIt \cite{anguera2007acoustic} - a weighted delay and sum beamformer was used for the conventional pipeline method. The reference channel selection in BeamformIt is performed using a metric named \textit{X-Corr} which is based on the average cross-correlation of one channel and all other channels.

Since the system can generalize to any number of channels, we only choose two channels while training to be memory efficient. All eight channels were used while testing. 
The batchsize was fixed as ``12" for all the experiments. 
The single channel E2E ASR baseline is trained by randomly choosing a channel when the batch consists of the REVERB data.

To regularize the ASR network when training the E2E frontends, we randomly choose whether to pass through the frontend  subnetworks or directly to the ASR encoder by also choosing a random channel. 
Also, while jointly training dereverberation-beamforming subnetworks, we randomly also skip the dereverberation part and give the input directly to the beamforming subnetwork.
We tried two types of masks for the beamforming subnetwork: a) Standard time-frequency (TF) mask b) One (time) dimensional mask like speech activity detection (SAD).
Dereverberation subnetwork always uses a TF mask. 

\begin{figure}[htbp]
  \centering
  \includegraphics[width=\linewidth]{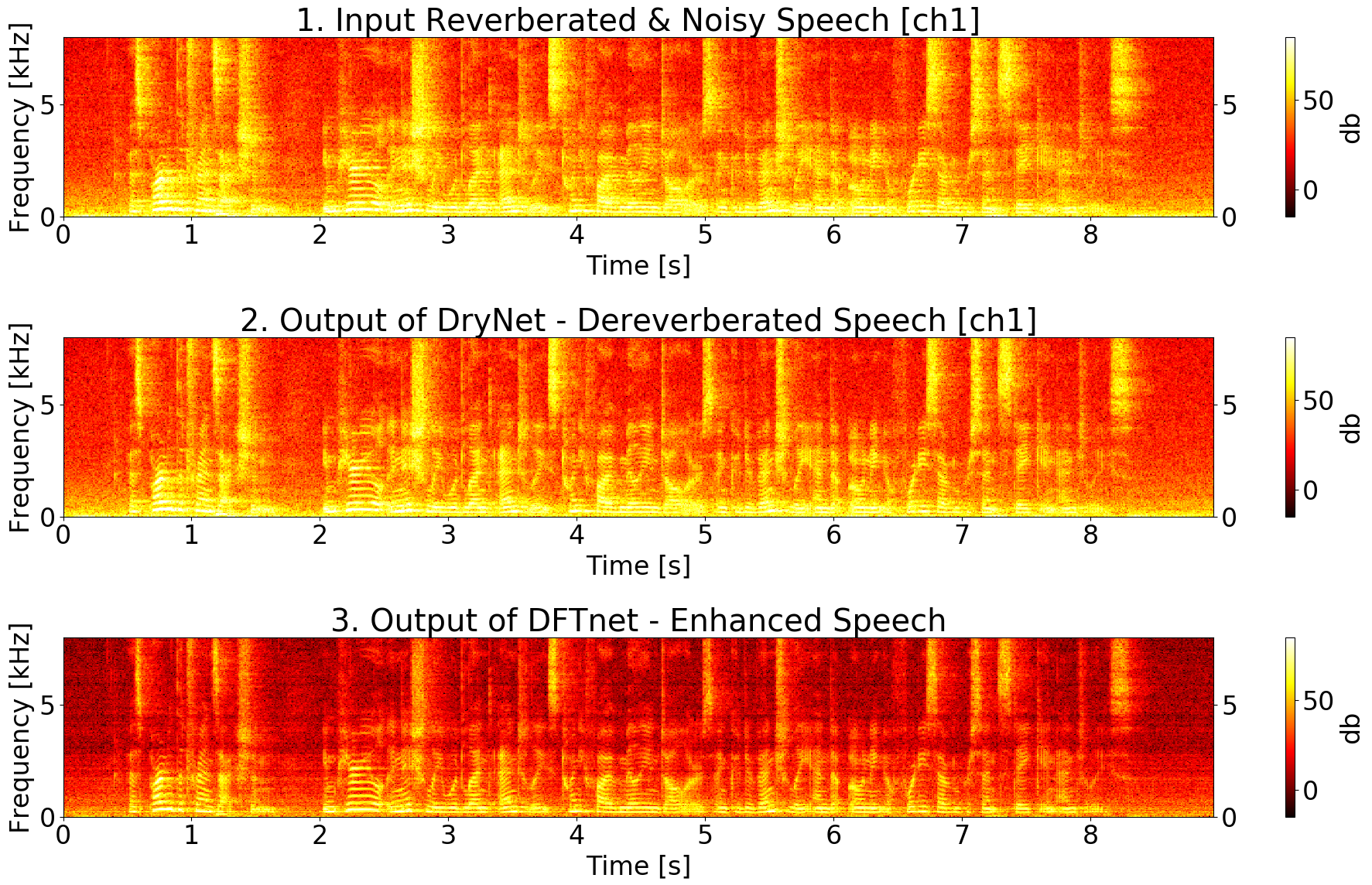}
  \caption{Sample real eval file - 1) Magnitude spectrum of the input reverberated noisy speech (ch1), 2) Output of dereverberation subnetwork after WPE filtering (ch1 magnitude), 3) The final enhanced signal output (magnitude) of frontend subnetworks after beamforming.}
  \label{fig:plot}
  \vspace*{-4mm}
\end{figure}

\subsection{Results \& discussion}
The ASR results comparing the performance of the pipeline frontend with the E2E frontend is given in Table~\ref{tab:WER}. 
The E2E baseline results are shown in the top row. 
It is compared with the results of the HMM-GMM baseline given as a part of the challenge \cite{kinoshita2016summary} and the E2E baseline performs better.
Table~\ref{tab:WER} shows that there is significant gain using dereverberation and/or beamforming in both pipeline and E2E frontends.
When we focus on the performance of solo E2E dereverberation or beamforming with ASR, they are better than their pipeline counterparts on some of the simulation and real near sets of REVERB, while there is a slight degradation on the real far sets of REVERB and DIRHA data.
This degradation could come from the mismatch in their conditions because the condition of the REVERB simulation training data is relatively less reverberant, and stationary compared with those of the REVERB real far and DIRHA data.  

The combination of E2E dereverberation and beamforming with TF mask gives the best results in most of the simulation test sets. 
The ``room1" simulated condition is relatively very clean compared to the other sets and we can infer that joint training does not distort the clean signal much when we use a TF mask. 
Finally, the combination of E2E dereverberation and beamforming with SAD mask along with attention based reference selection for the beamforming part is convincing overall on challenging real conditions. 
Interestingly this model gives 5.2\% relative improvement over the best pipeline method on the mismatched DIRHA set. 
This suggests that joint dereverberation and beamforming is well generalized even for more challenging mismatch data by considering the fact that it was trained only with the REVERB simulation data.

Finally, we compared the investigated method with the reference results using the same test set.
We chose \cite{tachioka2014dual} and \cite{alam2014use} from the submissions in the REVERB challenge official data track, which almost matches our training data conditions.
We also refer an end-to-end ASR system from \cite{wang2018stream}, which was trained with the DIRHA training data.
Joint dereverberation and beamforming with SAD mask attention based reference selection for the beamforming part often outperforms these results especially for the challenging REVERB real and DIRHA data conditions.



With these results, we can conclude that that it is possible to realize robust far-field ASR within an end-to-end manner by integrating dereverberation, beamforming, and ASR.
Note that this framework has extra benefits compared with the conventional pipeline methods as it is working without parallel data nor an iterative process.
Another benefit is that we can also generate enhanced signals from the output of each subnetwork, as discussed in Section \ref{sec:dftnet}. 
Interestingly, Figure \ref{fig:plot} shows that the dereverberated signal has less smearing in time and the effect of denoising is very clear in the final enhanced spectrogram from a clearer background.
This result indicates the investigated E2E multi-channel ASR system can perform dereverberation and denoising without any signal-level objective but with the ASR objective.

\section{Summary}
In this report, we investigated the ability of a recently developed multichannel end-to-end ASR model to work on noisy reverberant and mismatched environments. The investigated model jointly optimizes both beamforming and dereverberation components along with the ASR network only with the end-to-end ASR objective. 
We showed that this model is robust to mismatch conditions and gives comparable or better performance compared to existing pipeline methods.



\bibliographystyle{IEEEtran}

\bibliography{mybib}

\end{document}